\documentclass[%
 reprint,
superscriptaddress,
 amsmath,amssymb,
 aps,
pra,
longbibliography,
]{revtex4-1}

\usepackage{graphicx}% Include figure files
\usepackage{dcolumn}% Align table columns on decimal point
\usepackage{bm}% bold math
\usepackage{hyperref}% add hypertext capabilities
\usepackage{xcolor}
\usepackage{array}
\usepackage{braket}
\usepackage{multirow}
\begin{document}

\preprint{APS/123-QED}

\title{Origin of up-up-down-down magnetic order in Cu$_2$GeO$_4$}% Force line breaks with \\

\author{Danis I. Badrtdinov}
\affiliation{Theoretical Physics and Applied Mathematics Department, Ural Federal University, 620002 Yekaterinburg, Russia}

\author{Vladimir V. Mazurenko}
\affiliation{Theoretical Physics and Applied Mathematics Department, Ural Federal University, 620002 Yekaterinburg, Russia}

\author{Alexander A. Tsirlin}
\email{altsirlin@gmail.com}
\affiliation{Theoretical Physics and Applied Mathematics Department, Ural Federal University, 620002 Yekaterinburg, Russia}
\affiliation{Experimental Physics VI, Center for Electronic Correlations and Magnetism, Institute of Physics, University of Augsburg, 86135 Augsburg, Germany}

\date{\today}% It is always \today, today,

\begin{abstract}
We use density-functional band-structure calculations to explore the origin of the up-up-down-down (UUDD) magnetic order in Cu$_2$GeO$_4$ with the frustrated $J_1-J_2$ spin chains coupled into layers within the spinel-like crystal structure. In contrast to earlier studies, we find that the nearest-neighbor coupling $J_1$ should be negligibly small, owing to a nearly perfect compensation of the ferromagnetic direct exchange and antiferromagnetic superexchange. Under this condition, weak symmetric anisotropy of the exchange couplings gives rise to the UUDD order observed experimentally and also elucidates the non-trivial ordering pattern between the layers, whereas a small Dzyaloshinsky-Moriya interaction causes a spin canting that may generate local electric polarization. We argue that the buckling of the copper chains plays a crucial role in the suppression of $J_1$ in Cu$_2$GeO$_4$ and sets this compound apart from other $J_1-J_2$ chain magnets.
\end{abstract}

\maketitle

%\tableofcontents

\section{Introduction}
\label{sec:introduction}
Copper oxides built by chains of edge-sharing CuO$_4$ plaquettes serve as material prototypes of frustrated spin-$\frac12$ chains with competing nearest-neighbor and next-nearest-neighbor interactions $J_1$ and $J_2$, respectively. This simple spin model received ample attention~\cite{quantum-magnetism} triggered by the prospects of chiral, multipolar, and spin-nematic phases that may occur therein~\cite{kolezhuk2005,hikihara2008,sudan2009,meisner2009,zhitomirsky2010,sato2013,balents2016}. Whereas long-range order does not take place in one dimension (1D), interchain couplings in real materials will usually cause three-dimensional (3D) collinear or non-collinear order depending on the $J_2/J_1$ ratio. On the classical level, incommensurate spiral order appears for \mbox{$J_2/|J_1|>\frac14$}, whereas at $J_2/|J_1|<\frac14$ the second-neighbor coupling is not strong enough to tilt the spins, and the collinear ferromagnetic or up-down-up-down antiferromagnetic order form depending on the sign of $J_1$. Quantum effects preserve the spiral state in the case of ferromagnetic (FM) $J_1$~\cite{zinke2009}, but destroy the order and open a spin gap for antiferromagnetic (AFM) $J_1$ at $J_2/J_1>0.241$~\cite{white1996,eggert1996,furukawa2010}.

Real-world prototypes of the $J_1-J_2$ spin chains will typically follow one of these scenarios. The majority of quasi-1D copper oxides develop incommensurate spiral order~\cite{banks2009,capogna2010,zhao2012,willenberg2012}. Li$_2$CuO$_2$~\cite{boehm1998,lorenz2009}, Ca$_2$Y$_2$Cu$_5$O$_{10}$,~\cite{fong1999,kuzian2012}, and CuAs$_2$O$_4$~\cite{caslin2014} are notable exceptions, where $J_1$ is also FM, but spin alignment along the chains is purely ferromagnetic, owing to a smaller $J_2$.  Spin-chain compounds with AFM $J_1$ are more rare, although tentative indications of the spin-gap formation at $J_2/J_1>0.241$ have been reported~\cite{lebernegg2017}. 

One puzzling case in this series is Cu$_2$GeO$_4$~\cite{yamada2000} that reveals an unanticipated antiferromagnetic up-up-down-down (UUDD) order~\cite{zou2016} despite the prediction of FM $J_1$ and AFM $J_2$, both of the same magnitude~\cite{tsirlin2011}. This parameter regime would normally lead to the incommensurate spiral order, similar to LiCuVO$_4$, CuCl$_2$, and other $J_1-J_2$ cuprates. Here, we address this discrepancy and first analyze whether additional terms beyond $J_1$ and $J_2$ could destabilize the incommensurate order and give way to the UUDD state. This appears not to be the case, but instead $J_1$ is unusually weak in Cu$_2$GeO$_4$ and underlies the UUDD ground state of this compound.

The remainder of this paper is organized as follows. In Sec.~\ref{sec:str_el_properties}, we review the crystal structure of Cu$_2$GeO$_4$ and experimental information available for this material. Sec.~\ref{sec:methods} covers methodological aspects. In Sec.~\ref{sec:low_energy_model}, we estimate both isotropic and anisotropic exchange interactions in Cu$_2$GeO$_4$, and in Sec.~\ref{sec:Model solution} analyze the ensuing magnetic ground state. Ferromagnetic direct exchange appears to be crucial and merits further analysis presented in Sec.~\ref{sec:direct_exchange} followed by the analysis of experimental magnetic susceptibility in Sec.~\ref{sec:susceptibility} and a brief discussion and summary in Sec.~\ref{sec:discussion}.

%--------------------------------------------------------------------------------------------------------------
\section{Structure and properties of Cu$_2$GeO$_4$}
\label{sec:str_el_properties}
\begin{figure}
\includegraphics[width=0.49\textwidth]{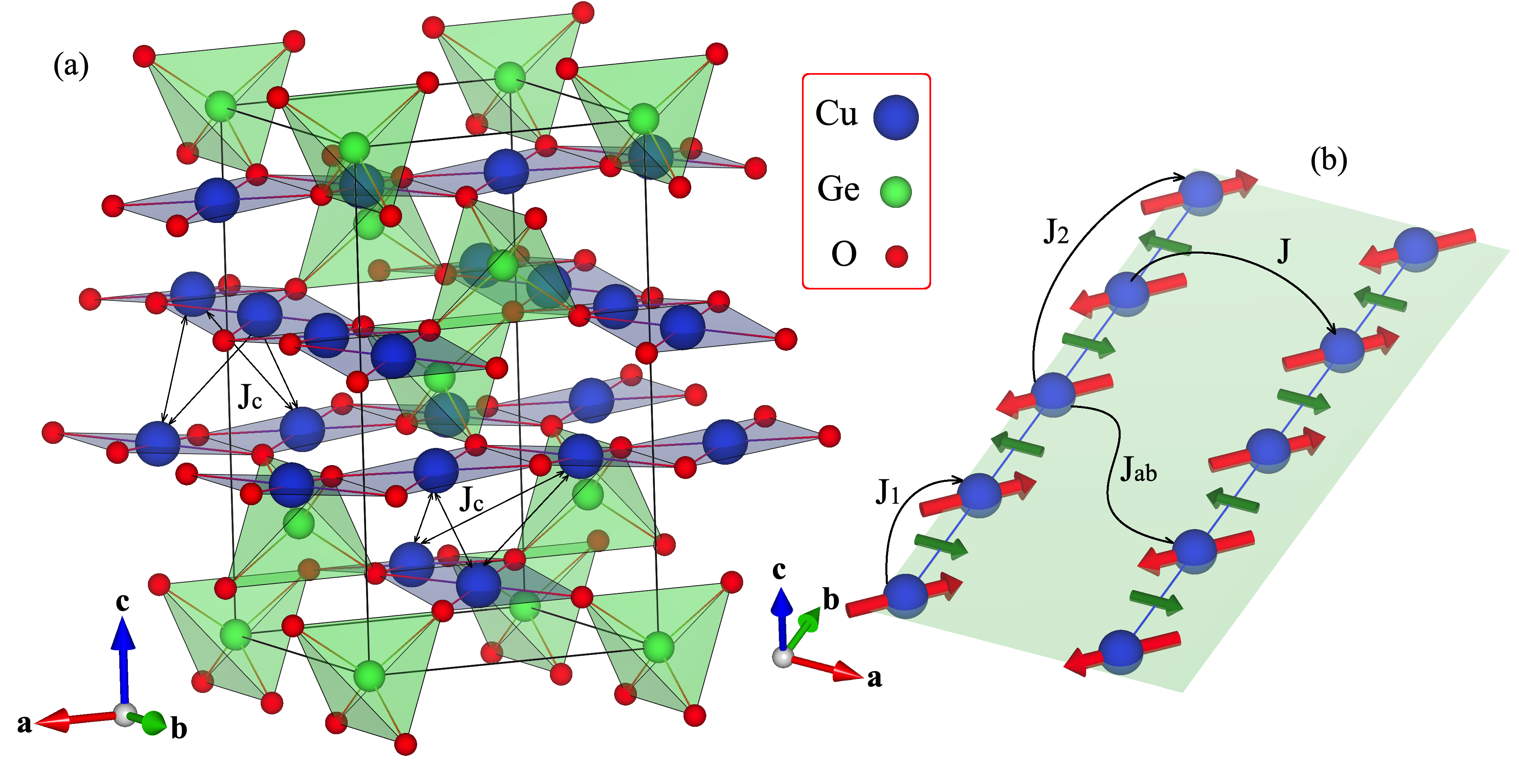}
\caption{(a) Crystal structure of Cu$_{2}$GeO$_{4}$. (b) Structural chains of copper atoms with the exchange integrals following the notation of Ref.~\onlinecite{tsirlin2011}. The green arrows show Dzyaloshinskii-Moriya vectors, while the red arrows represent electron spins that form the UUDD pattern according to Ref.~\onlinecite{zou2016}. Crystal structures were visualized using the VESTA software~\cite{vesta}. 
}
\label{fig:Crystal}
\end{figure}

Cu$_2$GeO$_4$ adopts a distorted spinel structure, where the Jahn-Teller effect inherent to Cu$^{2+}$ transforms CuO$_6$ octahedra into CuO$_4$ plaquettes~\cite{hegenbart1981}. The backbone of the structure is then formed by infinite chains of edge-shared plaquettes linked into a 3D network via the non-magnetic GeO$_4$ tetrahedra (Fig.~\ref{fig:Crystal}a). 

Magnetic susceptibility measurements revealed a broad maximum around 80\,K followed by an antiferromagnetic transition at $T_N\simeq 33$\,K~\cite{yamada2000}. This behavior is typical of low-dimensional and frustrated magnetism. In the case of Cu$_2$GeO$_4$, strong magnetic interactions are expected in the $ab$ plane, both along the chains of the Cu atoms ($J_1,J_2$) and perpendicular to the chains ($J$), see Fig.~\ref{fig:Crystal}b. The interactions $J_c$ between the planes are at least one order of magnitude weaker and form triangular loops together with $J_1$. This tentative magnetic model was confirmed by density-functional (DFT) band-structure calculations that yield $J_1\simeq -5.2$\,meV (FM) and $J_2\simeq 6.9$\,meV (AFM) as well as $J\simeq 11.2$\,meV. Even if the leading coupling $J$ runs perpendicular to the copper chains, magnetic order along these chains is still determined by the competition between $J_1$ and $J_2$, similar to the 1D $J_1-J_2$ model. Detailed numerical analysis confirmed the stability of the spiral order along the copper chains as well as the collinear spin arrangement perpendicular to the chains, where no significant frustration occurs~\cite{tsirlin2011}. 

Surprisingly, neutron diffraction data~\cite{zou2016} did not support this scenario and pinpointed the collinear UUDD order along the $J_1-J_2$ chains (Fig.~\ref{fig:Crystal}b). This spin configuration is uncommon for cuprates and has never been seen in the $J_1-J_2$ compounds before. Biquadratic exchange was considered as the driving force of this unusual order~\cite{zou2016} and may explain it indeed~\cite{kaplan2009}, but appears irrelevant to Cu$_2$GeO$_4$, because biquadratic terms do not exist for spin-$\frac12$ (they can be re-written as standard bilinear terms in the Hamiltonian~\cite{nagaev1982,mila2000}, see Appendix~\ref{app:Biquadratic}). Additionally, dielectric measurements revealed a clear anomaly in the permittivity at $T_N$, as well as a non-zero electric polarization that appears below $T_N$ in this formally centrosymmetric ($I4_1/amd$) crystal structure~\cite{zhao2018,yanda2018}. In the absence of spiral magnetic order that is typically associated with the electric polarization in chain cuprates~\cite{park2007,naito2007,seki2008,mourigal2011}, the origin of ferroelectricity in Cu$_2$GeO$_4$ remains controversial~\cite{zhao2018}. 

Here, we seek to throw some light on this problem from the \textit{ab initio} perspective. The conclusion of Ref.~\onlinecite{tsirlin2011} on the spiral order was based on the parametrization of an isotropic spin Hamiltonian, so it is natural to suspect, following Ref.~\onlinecite{zou2016}, that non-Heisenberg terms act against the spiral order and stabilize the UUDD one. We calculate such terms but find them to be small and affecting spin directions in the ordered state but not the nature of the ordered state itself. On the other hand, isotropic exchange couplings of Ref.~\onlinecite{tsirlin2011} have to be revised, eventually giving a clue to the formation of the UUDD order in Cu$_2$GeO$_4$. 

%--------------------------------------------------------------------------------------------------------------
\section{Methods}
\label{sec:methods}
In Ref.~\onlinecite{tsirlin2011}, the magnetic behavior of Cu$_2$GeO$_4$ was analyzed on the level of the Heisenberg spin Hamiltonian, 
\begin{eqnarray}
\hat {\mathcal{H}}^{\rm Heis} = \sum_{i>j} J_{ij} \hat {{\bf S}}_{i} \hat {{\bf S}}_{j}.
\label{eq:heis}
\end{eqnarray}
With $J_1\simeq -5.2$\,meV and $J_2\simeq 6.9$\,meV~\cite{tsirlin2011}, it leads to the spiral order along the copper chains at odds with the experiment. To account for the experimental UUDD order, additional terms may be invoked as follows,
\begin{eqnarray}
\hat {\mathcal{H}}^{\rm spin} = \hat{\mathcal H}^{\rm Heis}+ \sum_{i>j} {\bf D}_{ij}  [\hat {{\bf S}}_{i} \times  \hat {{\bf S}}_{j}] + \sum_{i>j} \hat {{\bf S}}_{i}  \tensor{\Gamma}_{ij} \hat {{\bf S}}_{j},
\label{eq:spin_ham}
\end{eqnarray}
where $\mathbf D_{ij}$ are Dzyaloshinskii-Moriya (DM) vectors and $\Gamma_{ij}$ are symmetric anisotropy tensors. The latter favor collinear spins and, therefore, may stabilize the UUDD order over the spiral one, whereas the former do not stabilize collinear spin configurations \textit{per se}, but may act against the spiral state. Specifically, in Cu$_2$GeO$_4$ the alternating directions of $\mathbf D_1$ (Fig.~\ref{fig:Crystal}b) are incompatible with the continuous spin rotation in the spiral. Biquadratic and other higher-order corrections do not appear as independent terms in the spin-$\frac12$ Hamiltonian~\cite{nagaev1982,mila2000}, see also Appendix~\ref{app:Biquadratic}.

Magnetic exchange parameters are obtained from DFT calculations performed within the generalized gradient approximation (GGA)~\cite{pbe96} implemented in Vienna ab initio Simulation Package ({\sc VASP})~\cite{vasp1,vasp2}. Additionally, the full-potential {\sc FPLO}~\cite{koepernik1999} and {\sc ELK}~\cite{ELK} codes were used. The crystal structure given in Ref.~\onlinecite{hegenbart1981} was employed in all calculations, similar to Ref.~\onlinecite{tsirlin2011}.

In the absence of electronic correlations, Cu$_2$GeO$_4$ features a metallic band structure with several bands crossing the Fermi level. The complex of four bands between $-0.6$ and 0.6\,eV corresponds to four Cu atoms in the primitive cell and arises from $d_{x^2-y^2}$ orbitals that are half-filled in Cu$^{2+}$. Electronic correlations split these bands and open a gap, in agreement with the insulating behavior of Cu$_2$GeO$_4$ expected from the green sample color~\cite{hegenbart1981,zhao2018}. The effect of correlations is modeled on the DFT+$U$+SO level, with all parameters of the spin Hamiltonian, Eq.~\eqref{eq:spin_ham}, extracted from total energies of ordered spin configurations using the mapping procedure~\cite{xiang2011}. Alternatively, we perform a model analysis based on hopping parameters of the uncorrelated band structure and additionally calculate ferromagnetic contribution to the exchange from the overlap of Wannier functions, as further explained in Sec.~\ref{sec:direct_exchange}. Similar methodology has been used in the previous DFT study~\cite{tsirlin2011}, but several technical details were different and proved to be crucial, as we show below.

The DFT+$U$+SO method relies on the parametrization of the Coulomb and Hund's exchange interactions $U_d$ and $J_H$, respectively.In Ref.~\onlinecite{tsirlin2011}, $U_d=6.5$\,eV and $J_H=1$\,eV were chosen empirically along with the around-mean-field (AMF) double-counting correction scheme that is more suitable for correlated metals. Here, we use the double-counting correction in the fully localized limit appropriate for insulators and obtain $U_d-J_H\sim 8.5$\,eV from the linear-response method~\cite{cococcioni2005}. Assuming $J_H=1$\,eV, we find $U_d=9.5$\,eV, which is similar to the parametrization that is typically used for copper oxides~\cite{janson2012,lebernegg2013,danil2017} in conjunction with the FLL flavor of the double-counting correction.   

Magnetic ground state of the spin Hamiltonian is obtained from the Luttinger-Tisza (LT) method considering spins as classical moments~\cite{luttinger1946,lyons1960}, 
\begin{eqnarray}
\mathcal{H}_{LT} = \sum_{\mathbf{k}} \mathbf{S}^{*}_{\mathbf{k}}  \tensor{J} (\mathbf{k} )  \mathbf{S}_{\mathbf{k}}, 
 \label{eq:LT}
\end{eqnarray}
where $\mathbf{S}_{\mathbf{k}}$ is the Fourier transform of the spin:
\begin{eqnarray}
\mathbf{S}_i = \frac{1}{\sqrt{N}} \sum_{\mathbf{k}} \mathbf{S}_{\mathbf{k}} e^{i \mathbf{k R}_i}
\end{eqnarray}

Diagonalization of Eq.~\eqref{eq:LT} yields~\cite{liu1026, sklan2013}
\begin{eqnarray}
\mathcal{H}_{LT} =  \sum_{\mathbf{k} \mu} \omega_{\mathbf{k} \mu} S^{*}_{\mathbf{k} \mu} S^{*}_{\mathbf{k} \mu},
\end{eqnarray}
where $S_{\mathbf{k} \mu} = \mathbf{S}_{\mathbf{k}} \hat{\mathbf{e}}_{\mathbf{k} \mu}$, $\omega_{\mathbf{k} \mu}$, and $\hat{\mathbf{e}}_{\mathbf{k} \mu}$ are corresponding eigenvalues and eigenvectors of $ \tensor{J} (\mathbf{k} )$.  The LT mode $S_{\mathbf{k} \mu}$ with the most negative eigenvalue $\omega_{\mathbf{k} \mu}$ is considered as an "optimal" mode with the wave vector $\mathbf{Q}_{\rm LT}$. If the constructed spin state $\{ \mathbf{S}_i \}$ is the linear combination of the optimal LT modes and complies with the "strong constraint" of $|\mathbf{S}_i|^2 = 1$, it can be considered as a ground state~\cite{sklan2013}.

We also calculate magnetic susceptibility of Cu$_2$GeO$_4$ using the \texttt{loop} algorithm~\cite{loop} of the ALPS simulation package~\cite{ALPS}. To this end, finite lattices with up to $16\times 16$ sites and periodic boundary conditions were used. 

\begin{figure}
\includegraphics[width=0.48\textwidth]{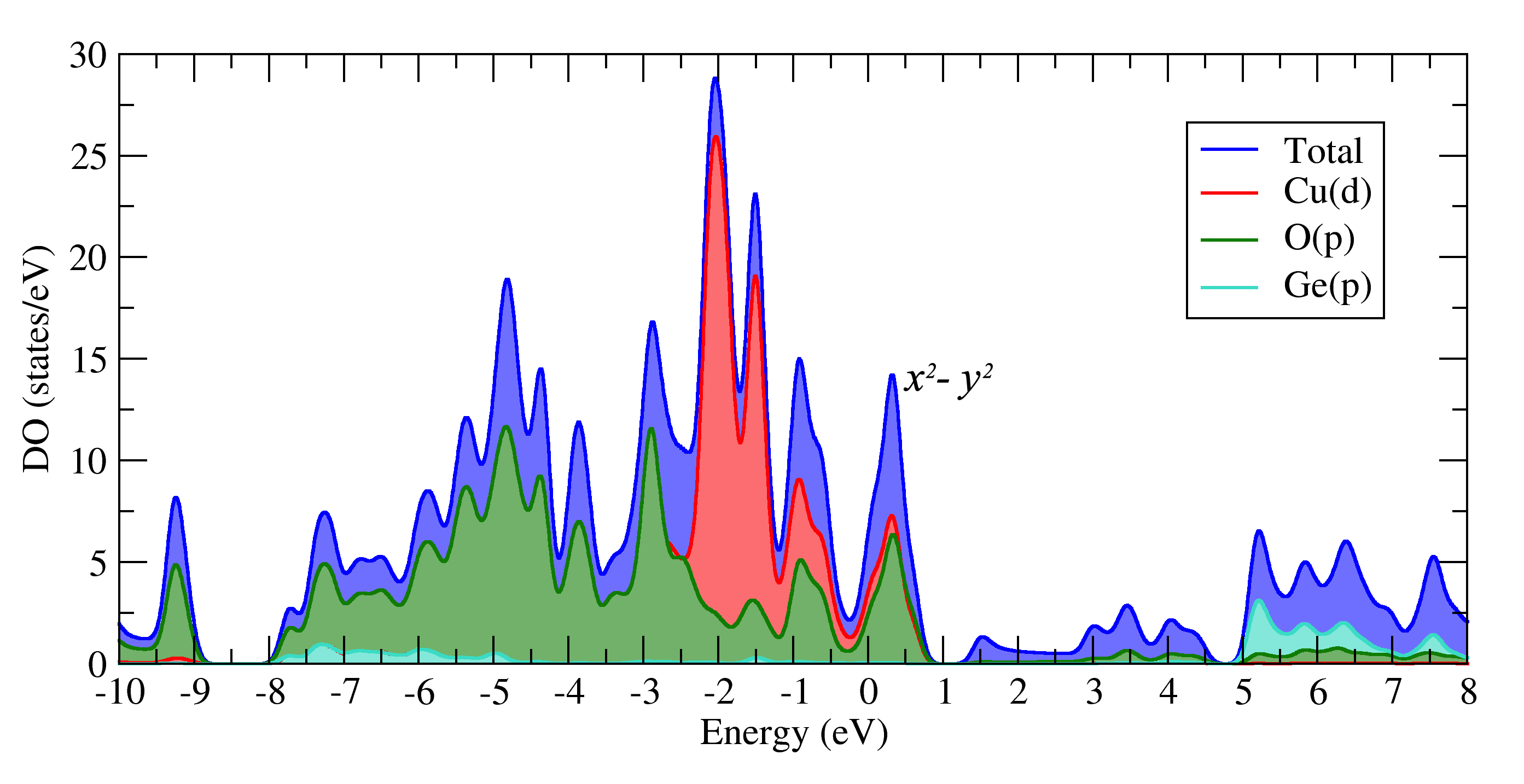}
\caption{ Electronic density of states for Cu$_{2}$GeO$_{4}$ obtained on the DFT (GGA) level.}
\label{fig:DOS}
\end{figure}

%--------------------------------------------------------------------------------------------------------------
\section{Results}
\label{sec:results}

\subsection{\label{sec:low_energy_model}Microscopic magnetic model}
Isotropic exchange couplings of the Heisenberg spin Hamiltonian, Eq.~\eqref{eq:heis}, are listed in Table~\ref{tab:Cu2GeO4_ISO}. DFT calculations were performed in three different codes that delivered largely consistent results for $J_2$ but not for $J_1$ that varies between $-0.2$\,meV in VASP and $-7.2$\,meV in FPLO, with ELK returning an intermediate value. This large spread of $J_1$ leads to a highly ambiguous physical picture, because the competition between $J_1$ and $J_2$ may be either strong (FPLO) or nearly non-existent (VASP). 

Other exchange couplings, including $J$ and $J_2$, are largely consistent between the different band-structure codes. This alone indicates that the ambiguity of $J_1$ does not stem from numerical inaccuracies, but reflects a complex nature of the coupling, which is short-range and combines dissimilar contributions of the direct exchange and superexchange, as opposed to the long-range couplings $J$ and $J_2$ dominated by the superexchange. To further exclude any technical issues related to the basis sets or energy convergence, we performed spin-polarized DFT calculations with $U_d=J_H=0$ and arrived at $J_1 \sim$ 40\,meV in all three codes~\footnote{The remaining spread of 3\,meV, which is less than 10\% of the absolute value, may be caused by the variation of magnetic moments between different spin configurations. This effect is unavoidable in uncorrelated calculations that underestimate local magnetic moment on Cu$^{2+}$.} (Table~\ref{tab:Energies}), thus confirming that the ambiguity of $J_1$ arises not from the different basis sets and not from the different treatment of the crystal potential, but from the way the DFT+$U$ correction is applied in each code. We also note that a variation of $U_d$ within the reasonable range of $1-2$\,eV does not improve the consistency between VASP and FPLO. 

\begin{table}
\centering
\caption [Bset]{ Isotropic exchange couplings (in meV) in Cu$_2$GeO$_4$ calculated within DFT+$U$ ($U_d=9.5$\,eV, $J_H=1$\,eV, FLL) using three different band-structure codes: FPLO, ELK and VASP. }
\label {basisset}
\begin{ruledtabular}
\begin {tabular}{l@{\hspace{1cm}}rr@{\hspace{0.5cm}}r}
   &  FPLO & ELK & VASP  \\
 \hline 
$J_{1}$    &   $-7.2$   & $-3.3$     &  $-0.2$          \\
$J_{2}$    &    6.0   &   5.0   &   5.6                         \\
$J$          &     9.0  &  7.8      &   8.5                        \\
$J_{ab}$  &    0.4   &  0.3     &   0.4                        \\
$J_{c}$    &   $-0.1$   &   0.5     &  $-0.1$              \\

\end{tabular}
\end{ruledtabular}
\label{tab:Cu2GeO4_ISO}
\end {table}

We also considered whether the mapping procedure could accidentally fail for $J_1$ if, for example, different spin configurations produced largely different local moments. However, this was not the case (Table~\ref{basisset}), and we are led to conclude that the problem with calculating $J_1$ is intrinsic. This coupling combines two contributions of different nature, the ferromagnetic direct exchange and antiferromagnetic superexchange, and different band-structure codes deliver very different estimates of these competing contributions.

We return to this $J_1$ problem in Sec.~\ref{sec:direct_exchange} but first consider anisotropic, non-Heisenberg terms that may also affect the ground state. These terms are obtained in VASP, because it delivers the most realistic estimate of $J_1$, as we show below. 

\begin{table}
\centering
\caption [Bset]{ DFT and DFT+$U$ ($U_d=9.5$\,eV, $J_d=1$\,eV, FLL) relative energies $E$ of magnetic configurations (in meV) used for evaluating $J_1 = ( E_{ \uparrow \uparrow } +  E_{ \downarrow \downarrow } - E_{ \downarrow \uparrow } -E_{ \uparrow \downarrow }) / 4S^2$ ($S = 1/2$) within three different band-structure codes. The $\uparrow$ and $ \downarrow$ symbols denote the magnetic moment alignment of two nearest-neighbour Cu$^{2+}$ ions coupled by the $J_1$ interaction, whereas magnetic moments of other Cu$^{2+}$ ions were fixed according to the procedure described in Ref.~\cite{xiang2011}. Magnetic moments of the two interacting copper sites, $m_1$ and $m_2$ (in $\mu_B$), are given for comparison. They are essentially similar between different spin configurations in DFT+$U$, but deviate from each other in spin-polarized DFT that underestimates local moments in correlated materials.  }
\label {basisset}
\begin{ruledtabular}
\begin {tabular}{llccc|ccc}
 &   &     \multicolumn{3}{c|}{DFT} &   \multicolumn{3}{c}{DFT+$U$} \\
  \hline
                                                   &        &   FPLO &    ELK   & VASP   &  FPLO & ELK       & VASP            \\
   \hline
   \multirow{3}{*}{$ \uparrow \uparrow $ } &   $E$ &  166.2  &  165.7  &  156.1  &   21.2   &  20.2    &   28.2            \\ 
    &  $m_1 $     &  0.65 &   0.62   &  0.55     &    0.80   &  0.80    &   0.80  \\
    &  $m_2 $    &  0.65 &   0.62    &  0.55    &    0.80  &  0.80     &   0.80  \\
    \hline
     \multirow{3}{*}{$  \downarrow \downarrow $ } &    $E$  &   0        &    0       &    0       &     0      &      0      &    0            \\ 
    &  $m_1 $     &  -0.57 &  -0.56    &  -0.52   &    -0.79  &  -0.79   &  -0.79    \\
    &  $m_2 $    &  -0.57 &   -0.56   &  -0.52    &    -0.79  & -0.79    &  -0.79  \\
     \hline
      \multirow{3}{*}{$\uparrow \downarrow $ } &    $E$  &   61.2   &   62.5   &  56.8  &    14.1   &  11.6     &   14.1            \\
    &  $m_1 $     &   0.64 &   0.62   &   0.56    &      0.80  &   0.80   &    0.80     \\
    &  $m_2 $    &  -0.55 &  -0.54   & -0.49    &    -0.79  &  -0.79   &   -0.79   \\
     \hline
     \multirow{3}{*}{$\downarrow \uparrow $ } &    $E$  &   63.7  &   64.7   &  58.9   &   14.3   &  11.9      &   14.3          \\
    &  $m_1 $     &  -0.55 &   -0.54   & -0.49    &     -0.79    &  -0.79   &  -0.79    \\
    &  $m_2 $    &   0.64 &    0.62     &  0.56     &       0.80  &   0.80    &    0.80    \\
     \hline    
    $J_1 $                       &                      &   41.3   &   38.5   &  40.4    & $-7.2$ &  $-3.3$ &  $-0.2$     \\
\end{tabular}
\end{ruledtabular}
\label{tab:Energies}
\end {table}

\begin{figure}[!h]
\includegraphics[width=0.48\textwidth]{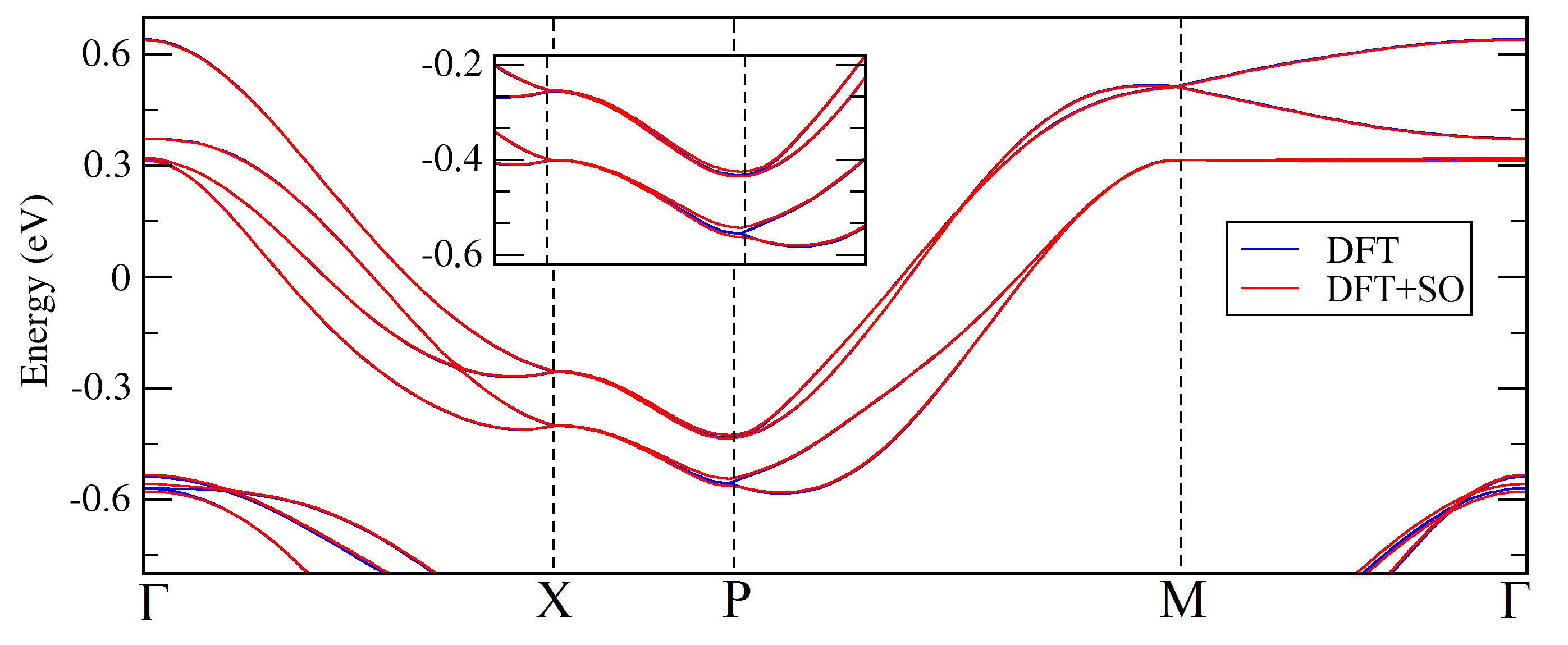}
\caption{Band structure of Cu$_2$GeO$_4$ near the Fermi level calculated within DFT and DFT+SO. The inset shows the magnified view to highlight the band splitting. }
\label{fig:SO}
\end{figure}

Anisotropic exchange is driven by the spin-orbit (SO) coupling. The effect of SO can be seen from the weak band splitting near the Fermi level at some of the high-symmetry $k$-points (Fig.~\ref{fig:SO}). The orbital moment of Cu$^{2+}$ reaches its highest value of 0.18\,$\mu_B$ for the direction perpendicular to the CuO$_4$ plaquettes, similar to other cuprates~\cite{danis2016, volkova2012}.

DM components for $J$ and $J_2$ are forbidden by the inversion symmetry. Therefore, the only non-vanishing DM vector is $\mathbf D_1$ that should lie in the $ab$ plane and perpendicular to the copper chains by virtue of the two mirror planes, one of them containing both Cu atoms and the other one passing through the middle of the Cu--Cu bond. From DFT+$U$+SO we find ${\bf D}_{01}$ = (0.01,~0,~0) meV for the plane with the copper chains running along the $b$ direction. In the neighboring planes with the Cu chains along $a$, the $\mathbf D_1$ vector has the same length but points along $b$ instead of $a$.

Symmetric components of the anisotropy for the interacting copper pairs shown in Fig.~\ref{fig:Crystal}b are similar in magnitude to the above DM vector (in meV),

\[\Gamma^{\mu \nu}_{J_1}= \left( \begin{array}{rrr} 
 -0.03 & 0.00  &  0.00  \\
   0.00 & 0.00  &  0.00  \\
   0.00 & 0.00  &  0.03  
 \end{array} \right). 
\]

\[\Gamma^{\mu \nu}_{J_2}= \left( \begin{array}{rrr} 
  0.00 &  0.00 &  0.00  \\
  0.00 &  0.00 &  0.02  \\
  0.00 &  0.02 &  0.00   
 \end{array} \right). 
\]

\[\Gamma^{\mu \nu}_{J}= \left( \begin{array}{rrr} 
-0.01 & 0.00 &  0.00  \\
 0.00 & 0.00 &  0.03  \\
 0.00 & 0.03 &  0.01   
\end{array} \right). 
\]

Here, the two-fold rotation axis along $c$ and the $bc$ mirror plane cancel all off-diagonal components of the $\tensor{\Gamma}_{J_1}$ tensor. In the case of $J_2$ and $J$, the rotation axis is missing, such that the nonzero $bc$ and $cb$ components become allowed. Taken together, these three tensors define the overall anisotropy matrix
\[\Gamma^{\mu \nu}_{\sum_J}= \left( \begin{array}{rrr} 
-0.08 & 0.00 &  0.00  \\
 0.00 & 0.00 &  0.10  \\
 0.00 & 0.10 &  0.08   
\end{array} \right).
\]
Its lowest eigenvalue defines $a$ as the easy direction for the layer with the copper chains running along $b$. Similar to $\mathbf D_1$, this easy direction changes to $b$ in the adjacent layers with the copper chains running along $a$. 

Although small compared to the isotropic exchange couplings, these anisotropic terms play crucial role in choosing spin directions in the magnetically ordered state (Sec.~\ref{sec:Model solution}). It is also worth noting that calculated anisotropic terms fulfill all symmetries of the system, and this fact lends credence to the DFT+$U$+SO results. Anisotropic interactions in cuprates are of superexchange nature~\cite{shekhtman1992,shekhtman1993} and thus easier to evaluate than $J_1$, which is strongly influenced by the direct exchange (Sec.~\ref{sec:direct_exchange}).

\subsection{\label{sec:Model solution}Model solution}  

\begin{figure}
\includegraphics[width=0.48\textwidth]{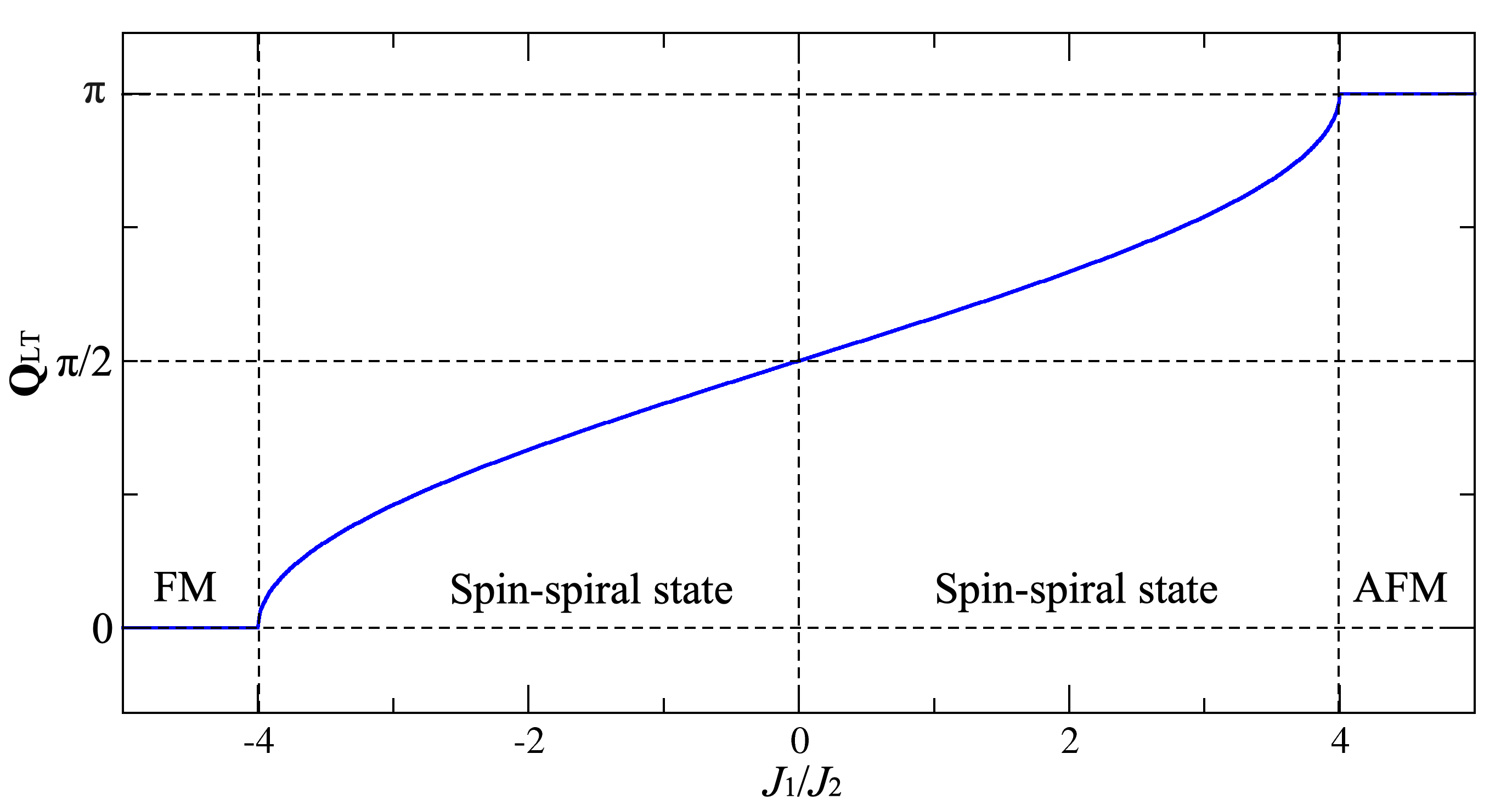}
\caption{ LT wave vector  $\mathbf{Q}^{LT}$  depending on the ratio  $J_1/J_2$ within isotropic $J_1 - J_2$ model. }
\label{fig:LT_Phase}
\end{figure}

We shall now use the LT method to determine the magnetic ground state. It is instructive to apply this method to the $J_1-J_2$ Heisenberg model first. Fig.~\ref{fig:LT_Phase} shows the LT wave vector depending on the $J_1/J_2$ ratio. The spiral order spans the region $-4<J_1/J_2<4$, as expected. However, at $J_1/J_2=0$ the wave vector $\mathbf Q_{\rm LT}=\pi/2$ corresponds to two possible solutions, the spin spiral with the pitch angle of $\pi/2$ (orthogonal spin configuration) and the collinear UUDD order that has been observed in Cu$_2$GeO$_4$ experimentally~\cite{zou2016}. We may thus expect the UUDD order at $J_1=0$. 

\begin{figure}
\includegraphics[width=0.48\textwidth]{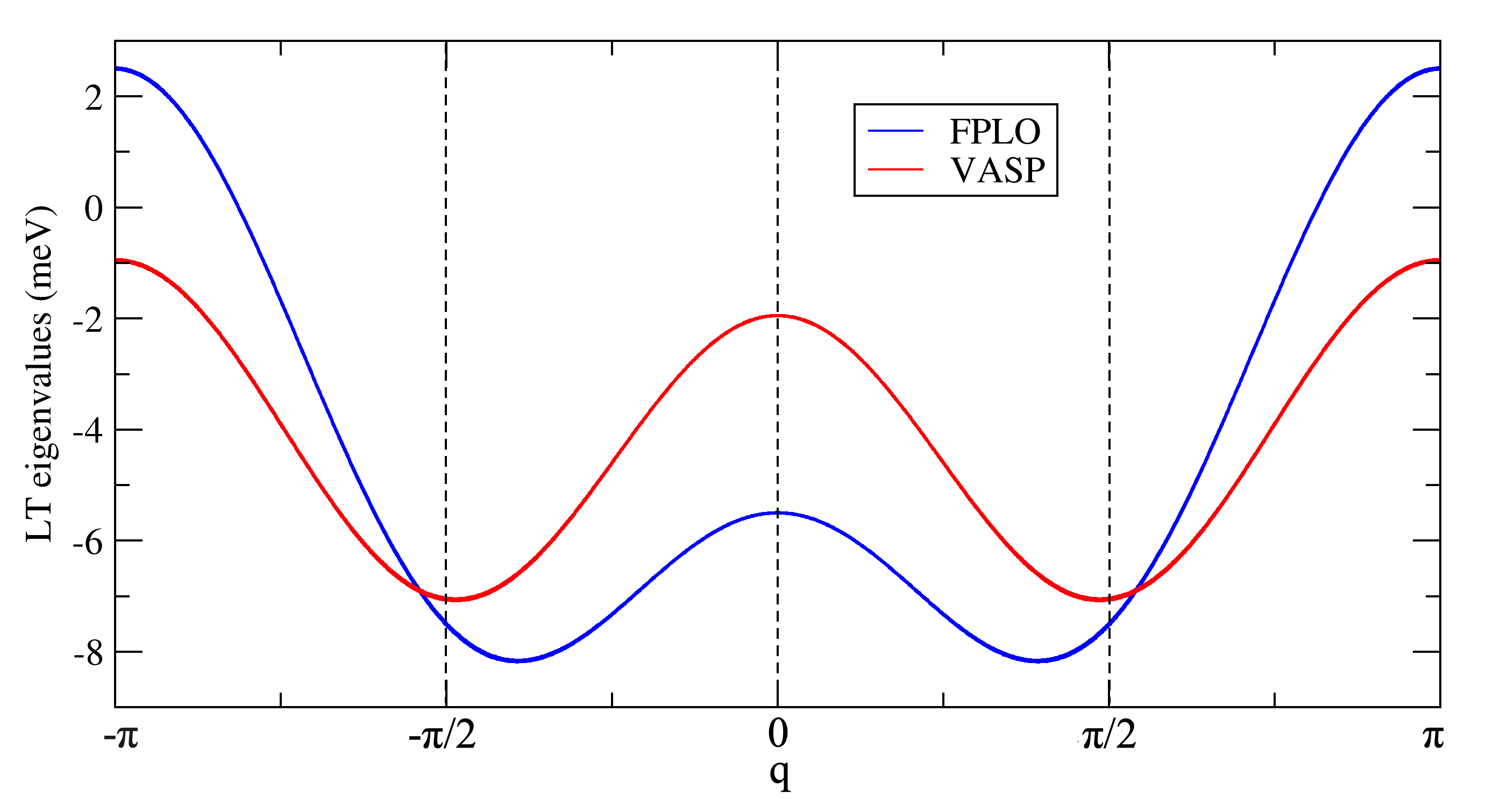}
\caption{Luttinger-Tisza eigenvalues along ${\bf q} = (\frac{\pi}{2}, q)$ obtained using the $J_{ij}$ values from the {\sc FPLO} and {\sc VASP} codes, respectively (Table~\ref{basisset}). }
\label{fig:LT}
\end{figure}

Fig.~\ref{fig:LT} shows the LT wavevectors obtained for the full set of in-plane exchange couplings ($J_1$, $J_2$, $J$, and $J_{ab}$) from {\sc FPLO} and {\sc VASP}. The {\sc FPLO} results clearly lead to the incommensurate position of the minimum and stabilize the spiral order. Weak anisotropic terms presented above do not change the $\mathbf q$-vector significantly. On the other hand, the {\sc VASP} results produce the minimum at $\mathbf q=(\pi/2,\pi/2)$ compatible with the UUDD order or with the spin spiral having the $\pi/2$ rotation along the copper chain. The former state is collinear and thus benefits from the symmetric anisotropy, the $xx$-term of $\Gamma^{\mu \nu}_{\sum_J}$. The spiral state will, on the other hand, gain less energy from $\Gamma^{\mu \nu}_{\sum_J}$, because different spin directions are present.  This has been confirmed by a direct DFT+$U$+SO calculation for the UUDD state and for the spin spiral with the pitch angle of $\pi/2$. The UUDD state is lower in energy by 0.45 meV per Cu$^{2+}$ ion. 

We conclude that the UUDD order can be competitive with the spiral order, but around $J_1=0$ only. As soon as this condition is fulfilled, symmetric anisotropy present in Cu$_2$GeO$_4$ favors the UUDD order over the spiral one, because all spins can follow easy direction in the UUDD state but not in the spiral. This way, the negligibly small $J_1$ is the necessary condition for the formation of the UUDD order. We shall further justify this condition in Sec.~\ref{sec:direct_exchange} below, but first demonstrate that the combination of $J_1=0$ and weak symmetric anisotropy explains not only the UUDD order along the copper chains, but also all other features of the experimental magnetic structure. 

\begin{figure}
\includegraphics[width=0.48\textwidth]{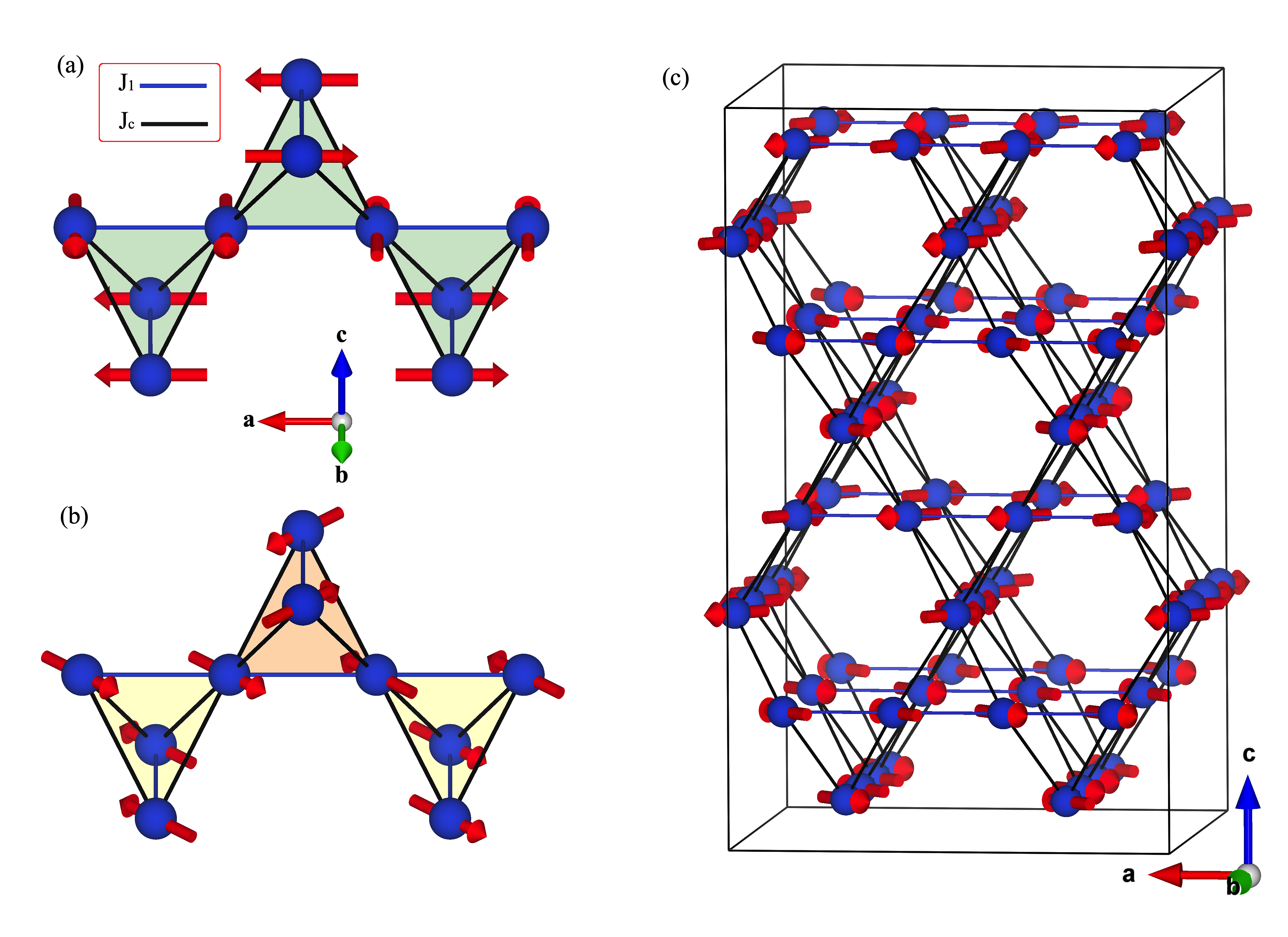}
\caption{ Magnetic order in the $J_2-J$ model stabilized under (a) the action of anisotropic terms $\Gamma$; (b) the combination of anisotropic terms $\Gamma$ and antiferromagnetic interlayer coupling $J_c$. The ordering pattern in (b) is identical to the experimental magnetic structure of Cu$_2$GeO$_4$, depicted in (c)~\cite{zou2016}. }
\label{fig:Octahedra}
\end{figure}

In the absence of the interlayer coupling, the symmetric anisotropy $\Gamma^{\mu \nu}_{\sum_J}$ puts spins along $b$ in the layer where the copper chains run along $a$, and along $a$ in the layer where the copper chains run along $b$. This would lead to orthogonal spin directions in the neighboring layers (Fig.~\ref{fig:Octahedra}a) and becomes compatible with the scenario of frustrated interlayer couplings $J_c$. However, the UUDD order releases the frustration on those tetrahedra, where spins are parallel along the $J_1$ bonds, and such tetrahedra may gain energy from $J_c$. Then $\mathbf a+\mathbf b$ or $\mathbf a-\mathbf b$ are chosen as compromise spin directions between the two layers (Fig.~\ref{fig:Octahedra}b). In contrast, the tetrahedra with antiparallel spins along $J_1$ remain frustrated and enjoy the orthogonal spin arrangement (Fig.~\ref{fig:Octahedra}b). This leads to the peculiar magnetic order observed in Cu$_2$GeO$_4$. The spin direction alternates between $\mathbf a+\mathbf b$ and $\mathbf a-\mathbf b$ in every second layer in response to the frustration present on one half of the Cu$_4$ tetrahedra and absent on the other half.

\begin{figure}
\includegraphics[width=0.48\textwidth]{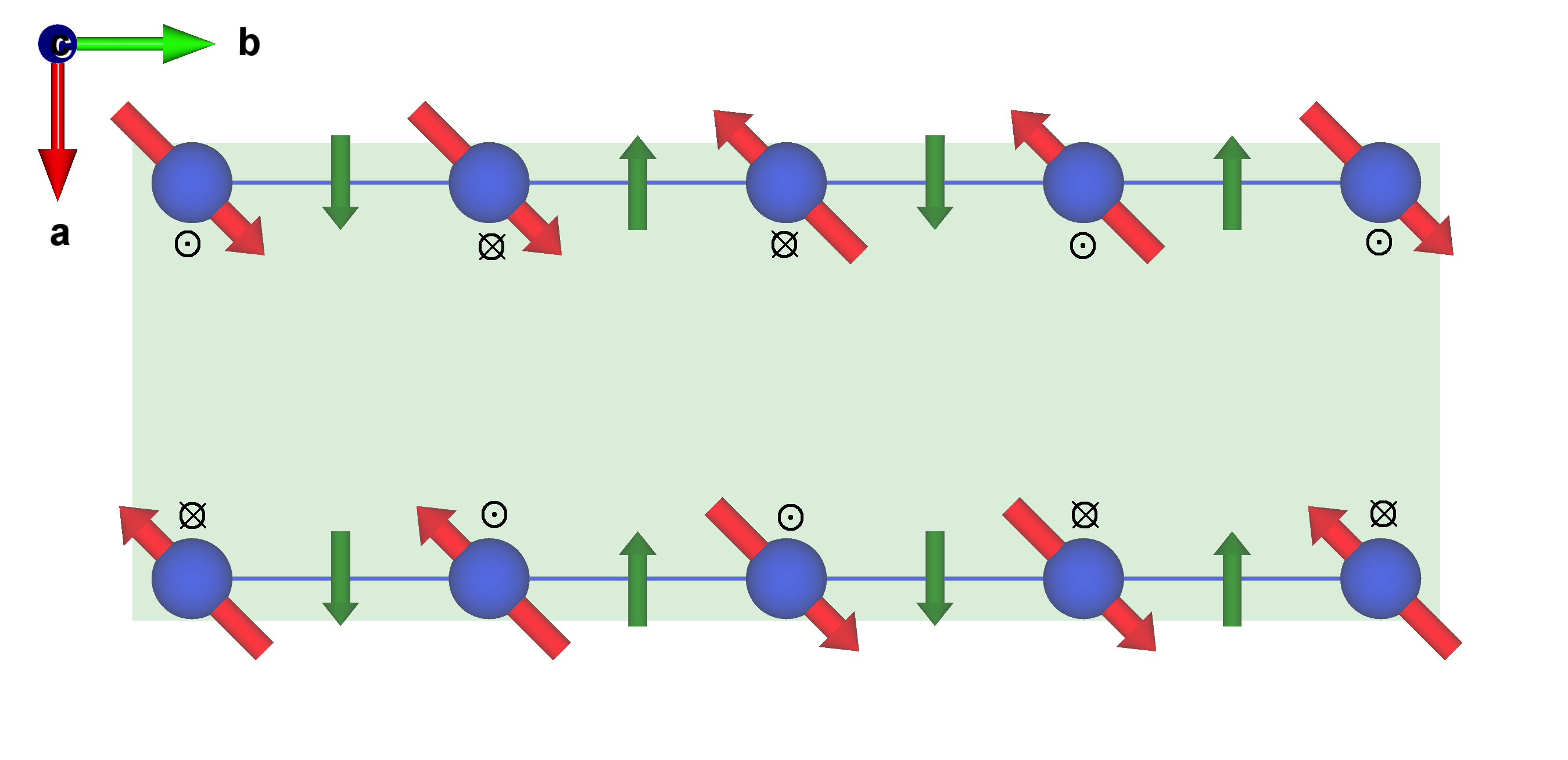}
\caption{ Small deviation from the collinear UUDD order caused by the nearest-neighbor DM interactions in Cu$_2$GeO$_4$. The green arrows represent the DM vectors, while the red arrows depict the spins of the magnetic Cu$^{2+}$ ions forming the UUDD pattern. Black circles schematically show the direction of spin canting along $c$.}
\label{fig:Disortion}
\end{figure}

The DM interactions were not considered so far, because they neither stabilize nor destabilize the collinear UUDD state. They may, however, introduce weak spin canting as shown in Fig.~\ref{fig:Disortion}. This canting is fully compensated within each chain and does not produce any net magnetic moment. From the $\mathbf D_1$ value obtained in Sec.~\ref{sec:low_energy_model} and from the direct relaxation of the magnetic structure within {\sc VASP}, we estimate only a weak non-collinearity with the canted moment of about 0.005\,$\mu_B$. Such a moment is clearly too small to be detected by powder neutron diffraction~\cite{zou2016}, but is allowed by symmetry and may be relevant to the development of electric polarization, as we further explain in Sec.~\ref{sec:discussion}.

%--------------------------------------------------------------------------------------------------------------
\subsection{\label{sec:direct_exchange}Direct exchange}
Having established $J_1=0$ as the necessary condition for the UUDD order, we now discuss the microscopic origin of $J_1$ and the reasons for the full compensation of this coupling in Cu$_2$GeO$_4$. Magnetic couplings in insulators are generally composed of two contributions, the kinetic term due to the superexchange, $J^{\rm kin}_{ij} = 4 t^2_{ij} /  \widetilde{U}_{ij}$, and the potential term due to the direct exchange interaction $J^{F}_{ij}$ arising from the direct overlap of the magnetic orbitals~\cite{anderson1959}. We then write an isotropic exchange coupling in the form
\begin{equation}
J_{ij} = \frac{4 t^2_{ij}}{\widetilde{U}_{ij}}+ 2J^{F}_{ij},
\end{equation}
where $t_{ij}$ is the hopping integral, $\widetilde{U}_{ij} = U_{ii} - U_{ij}$ is an effective screened Coulomb repulsion parameter~\cite{mazurenko2007,danis2018}, and $J^F_{ij}$ is the direct exchange.  

The direct exchange depends on the overlap between the magnetic orbitals. This overlap is very sensitive to hybridization effects, because spin polarization spreads onto ligands, which contribute to the overlap and largely determine the $J^{F}_{ij}$ values in real materials. This hybridization effect can be captured using Wannier functions that serve as a realistic representation of the magnetic orbitals. Here, we use maximally localized Wannier functions for Cu$^{2+}$~\cite{marzari1997} and illustrate the role of the hybridization by calculating three-dimensional magnetic form-factors $F(\mathbf{q})$ as Fourier transforms of the Wannier orbitals~\cite{mazurenko2015},
\begin{equation}
 F(\mathbf{q}) = \int |W(\mathbf{r})|^{2} e^{-i \mathbf{q r}}  d \mathbf{r}.
\end{equation}

\begin{figure}
\includegraphics[width=0.48\textwidth]{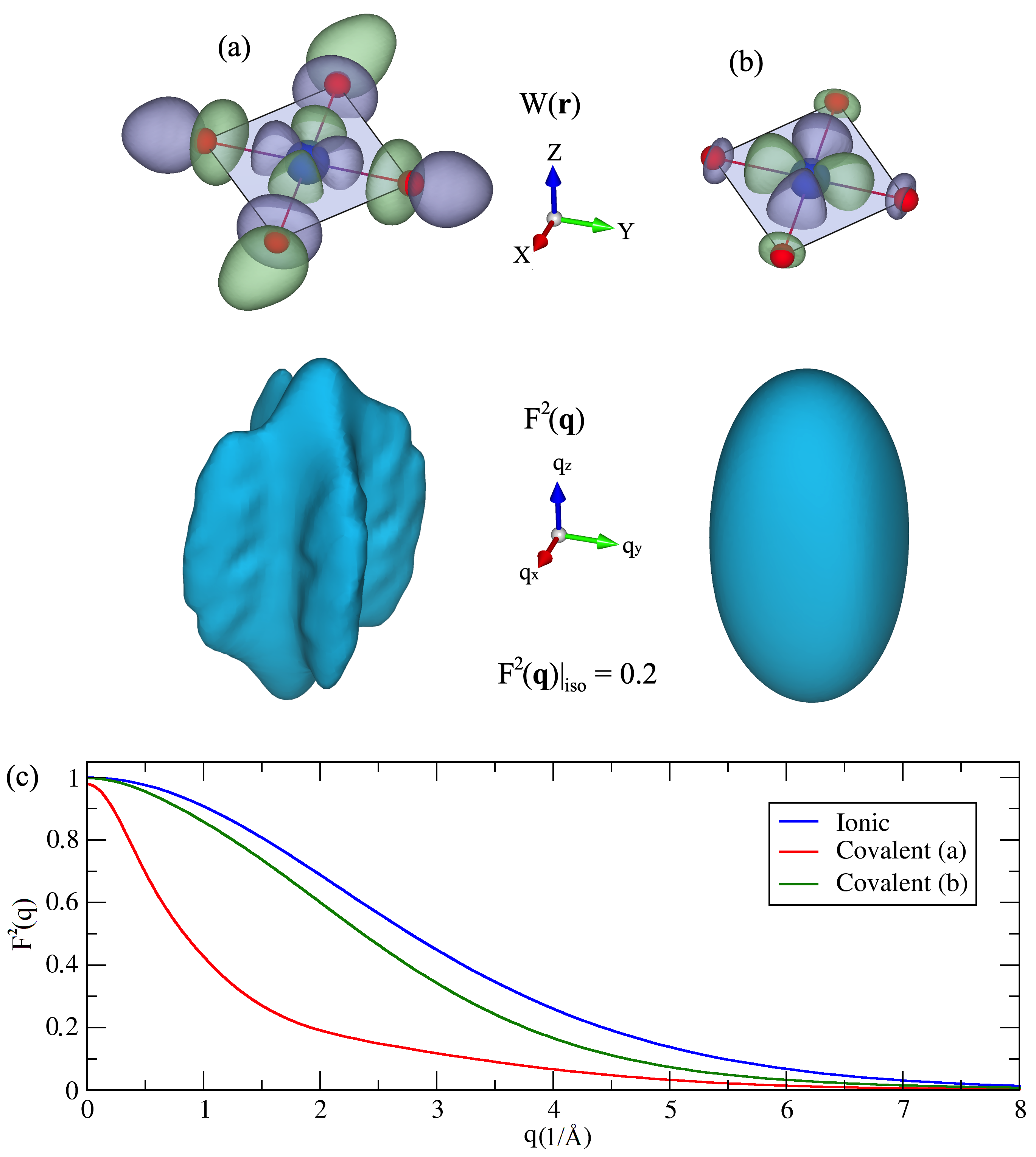}
\caption{ The Wannier orbitals of $x^2-y^2$ symmetry obtained  (a) within the one-band model that includes the states at the Fermi level only, and (b) by taking all Cu $d$ and oxygen $p$ states into account (Fig.~\ref{fig:DOS}). Different colors denote different phases of the Wannier orbital. The lower graphs are three-dimensional magnetic form-factors represented on the same isosurface level.  (c) A comparison between the ionic Cu$^{2+}$ form-factor obtained within the 3-Gaussian approximation~\cite{PJBrown} and the covalent form-factors calculated by powder-averaging of (a) and (b).  }
\label{fig:FF}
\end{figure}

In Fig.~\ref{fig:FF}, we compare two scenarios: i) Wannier functions calculated for four Cu $d_{x^2-y^2}$ bands in the vicinity of the Fermi level; ii) Wannier functions calculated for all Cu $3d$ and O $2p$ states. The second case leads to the lower oxygen contribution and renders $F(\mathbf q)$ more symmetric, similar to the purely ionic form-factor for Cu$^{2+}$. In contrast, case i) captures the full Cu--O hybridization, makes $F(\mathbf q)$ less symmetric, and causes a faster decay at higher $q$'s. Similar effects were reported for other Cu$^{2+}$ oxides~\cite{mazurenko2015,walter2009} and may be responsible for the reduced ordered moment of $0.89(5)$\,$\mu_B$ determined by neutron diffraction using the ionic form-factor for Cu$^{2+}$~\cite{zou2016}. 

We are now in a position to calculate bare Coulomb (both on-site $V_{ii}$ and intersite $V_{ij}$) and non-local direct exchange $J^{F}_{ij}$ integrals   in the basis of Wannier functions $W_i(\boldsymbol{r})$,
\begin{equation}
V_{ij} = \int  \frac{ W^{*}_i(\boldsymbol{r}) W_i(\boldsymbol{r})  W^{*}_j(\boldsymbol{r}') W_j(\boldsymbol{r}') }{|\boldsymbol{r}-\boldsymbol{r}'|}  d\boldsymbol{r} d\boldsymbol{r}',
\label{Coulomb_bare}
\end{equation}
and
\begin{equation}
J^{F}_{ij} = \int  \frac{ W^{*}_i(\boldsymbol{r}) W_j(\boldsymbol{r})  W^{*}_j(\boldsymbol{r}') W_i(\boldsymbol{r}') }{|\boldsymbol{r}-\boldsymbol{r}'|}  d\boldsymbol{r} d\boldsymbol{r}'.
\label{Coulomb_bare}
\end{equation}

Screening effects were captured on the level of random phase approximation (RPA)~\cite{RPA}. However, energy bands at the Fermi level are also involved in the screening processes and cause a "self-screening" that needs to be excluded in order to evaluate the partially screened, realistic Coulomb parameters~\cite{cRPA}. Therefore, we utilized constrained RPA and obtained the on-site $U_{ii}$ and intersite $U_{ij}$ Coulomb parameters listed in Table~\ref{tab:Cu2GeO4_RPA}. As for the non-local direct exchange, its evaluation within constrained RPA requires a very accurate integration within the Brillouin zone and proved to be unfeasible. Therefore, we used standard RPA and calculated the fully-screened $J^{F}_{\rm scr}$ that gives the lower bound for the FM contribution and also allows for a comparison between different exchange pathways and different compounds.

\begin{table}
\centering
\caption [Bset]{ The magnetic model parameters for Cu$_2$GeO$_4$, Li$_2$CuO$_2$, and CuGeO$_3$ in the basis of Wannier functions (in meV). The corresponding Coulomb parameters calculated by using random phase approximation (RPA). In particular, self-screening effects were extracted from Coulomb integrals (constrained RPA), whereas the direct exchange integrals were evaluated without excluding the self-screening effects. We use the notation $\widetilde{U}_{ij} = U_{ii} - U_{ij}$ for the effective Coulomb parameter. }
\begin{ruledtabular}
\setlength{\extrarowheight}{5pt}
\begin {tabular}{l|ccc|cc|cc}
             & \multicolumn{3}{c|}{Cu$_2$GeO$_4$} &  \multicolumn{2}{c|}{Li$_2$CuO$_2$}   & \multicolumn{2}{c}{CuGeO$_3$}  \\
             & \multicolumn{3}{c|}{$U_{ii}$ = 2\,eV} &  \multicolumn{2}{c|}{$U_{ii}$ = 3.6\,eV}   & \multicolumn{2}{c}{$U_{ii}$ =  4.1\,eV}  \\
              \hline 
             &   $J_{1}$ &  $J_{2}$ &   $J$  &   $J_{1}$  &  $J_{2}$  & $J_{1}$  & $J_{2}$ \\
              \hline
 $t_{ij}$  &    119    &    86  &   121   &  70 & 81 & 205 & 76  \\
 $U_{ij}$ &   500   &   200 &   330   &  1510  &  680  & 1480 & 730    \\
 $ 4t^2_{ij}/ \widetilde{U}_{ij}$ &  37.8 &  16.4 &   35.1    &  9.4   & 9.0  & 64.2 & 6.9   \\
 $2J^{F}_{\rm bare}$ &  $-62.4$ &   $-15.0$ &  $-35.0$  & $-88.6$  & $-6.8$  & $-72.4$ & $-9.0$  \\
 $2J^{F}_{\rm scr}$   &  $-30.2$  &   $-2.8$  &  $-5.6$   & $-43.0 $ & $-2.4 $  & $-34.4$ & $-2.6$  \\  
   \hline
  $J^{\rm VASP}_{ij}$ &  $-0.2$    &    5.6  &    8.5  & $-11.6 $ & 5.5 &  18.7  & 3.7  \\

\end {tabular}
\end{ruledtabular}
\label{tab:Cu2GeO4_RPA}
\end {table}

Wannier representation of the band structure also gives access to the hopping integrals $t_{ij}$. This way, we obtain both FM and AFM contributions to the exchange couplings in Cu$_2$GeO$_4$, as listed in Table~\ref{tab:Cu2GeO4_RPA}. In the case of $J_2$ and $J$, AFM superexchange clearly dominates over the fully screened direct exchange, and the overall AFM couplings ensue. On the other hand, $J^F_{\rm scr}$ for $J_1$ is only slightly smaller in magnitude than $J^{\rm kin}$, suggesting that $J_1$ may be close to zero, as {\sc VASP} DFT+$U$ calculations predict.

It is also instructive to juxtapose Cu$_2$GeO$_4$ with other compounds containing copper chains. To this end, we choose Li$_2$CuO$_2$ with its large FM $J_1\simeq -19.6$\,meV~\cite{lorenz2009} and CuGeO$_3$ where $J_1$ was proposed to be AFM~\cite{castilla1995}. The main structural difference between these compounds lies in the nearest-neighbor Cu--O--Cu angle that increases from $93.97^{\circ}$ in Li$_2$CuO$_2$ (at 1.5\,K)~\cite{sapina1990} to $99.24^{\circ}$ in CuGeO$_3$ (at 20\,K)~\cite{braden1996}. The nearest-neighbor hopping is, consequently, enhanced and makes $4t_1^2/\tilde U_{ij}$ much larger than $2J^F_{\rm scr}$ (Table~\ref{tab:Cu2GeO4_RPA}). This way, the crossover from FM $J_1$ in Li$_2$CuO$_2$ to AFM $J_1$ in CuGeO$_3$ is caused by the increased Cu--O--Cu angle, while all other microscopic parameters of these compounds are similar.  

Coming now to Cu$_2$GeO$_4$, we realize that its hopping integral $t_1$ and, thus, the AFM contribution to $J_1$ are intermediate between those of Li$_2$CuO$_2$ and CuGeO$_3$. The FM contribution $2J^F_{\rm scr}$ is, on the other hand, reduced in magnitude. Both aspects create suitable conditions for the cancellation of the FM and AFM contributions, leading eventually to $J_1\simeq 0$. Such an unusual behavior may be rooted in peculiarities of the Cu$_2$GeO$_4$ crystal structure. The nearest-neighbor Cu--O--Cu angle of $91.57^{\circ}$~\cite{zhao2018} is in fact lower than in Li$_2$CuO$_2$, so naively one would expect an even lower $t_1$, which is not the case. Weak buckling of the copper chains (Fig.~\ref{fig:Crystal}a) appears to be crucial here, because it reduces the direct overlap responsible for $J^F_{\rm scr}$ and, simultaneously, shortens the nearest-neighbor Cu--Cu distance from 2.860\,\r A in Li$_2$CuO$_2$~\cite{sapina1990} to 2.796\,\r A in Cu$_2$GeO$_4$~\cite{zhao2018}, thus enhancing the direct $d-d$ hopping. A similar argument can be applied to linarite, PbCu(SO$_4$)(OH)$_2$, that also features buckled copper chains with an intermediate Cu--Cu distance of 2.823\,\r A. Indeed, its $J_1\simeq -8.6$\,K~\cite{rule2017} or $-13.8$\,K~\cite{cemal2018} is smaller in magnitude than in Li$_2$CuO$_2$, but still on the ferromagnetic side.

%--------------------------------------------------------------------------------------------------------------
\subsection{Magnetic susceptibility}
\label{sec:susceptibility}

Further support for the $J_1\simeq 0$ scenario can be garnered by analyzing magnetic susceptibility of Cu$_2$GeO$_4$. According to Table~\ref{tab:Cu2GeO4_ISO}, the minimal magnetic model for this compound should only include the coupling $J$ that forms spin chains perpendicular to the structural chains of the Cu atoms. Adding the coupling $J_2$ connects these spin chains into a rectangular spin lattice. We used both models to fit the experimental susceptibility data from Ref.~\onlinecite{yamada2000}.

\begin{figure}
\includegraphics[width=0.48\textwidth]{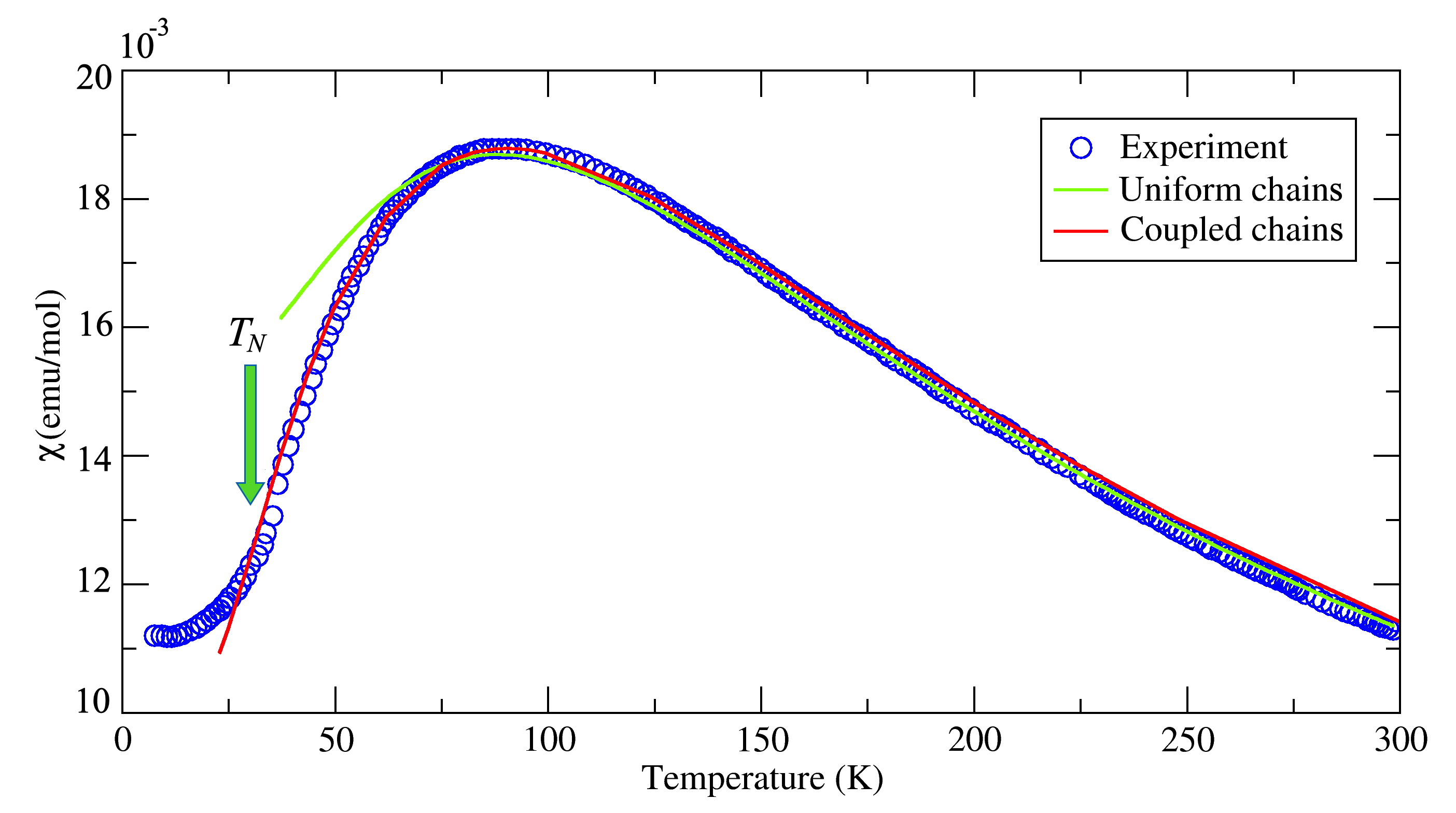}
\caption{Experimental magnetic susceptibility of Cu$_2$GeO$_4$~\cite{yamada2000} and its fits with the spin-chain ($J$-only) and rectangular-lattice ($J-J_2$) models.   }
\label{fig:CHI}
\end{figure} 
The spin-chain model leads to a decent fit above 100\,K with $J=12.1$\,meV, $g=2.23$, and the temperature-independent term $\chi_0=-7.7\times 10^{-5}$\,emu/mol, but at lower temperatures this model overestimates the experimental susceptibility (Fig.~\ref{fig:CHI}), suggesting that antiferromagnetic interchain couplings are at play. By including $J_2$, we obtain an excellent fit down to $T_N$ with $J=10.7$\,meV, $J_2=5.3$\,meV, $g=2.33$, and $\chi_0=-0.0001$\,emu/mol (Fig.~\ref{fig:CHI}). The fitted values of $J$ and $J_2$ are in good agreement with the DFT estimates in Table~\ref{tab:Cu2GeO4_ISO}. Moreover, we confirm that the experimental susceptibility of Cu$_2$GeO$_4$ is compatible with the $J_1\simeq 0$ scenario.

%--------------------------------------------------------------------------------------------------------------
\section{Discussion and summary}
\label{sec:discussion}
We have shown that the UUDD magnetic structure of Cu$_2$GeO$_4$ can be obtained in the limit of the weak nearest-neighbor coupling $J_1\simeq 0$. Two additional ingredients, frustration on half of the Cu$_4$ tetrahedra and orthogonal easy directions in the adjacent copper layers, explain all peculiarities of the experimental magnetic structure with its two spin directions, $\mathbf a\pm\mathbf b$, that change in every second layer~\cite{zou2016}. We revise the previous \textit{ab initio} results~\cite{tsirlin2011} and establish the new magnetic model of Cu$_2$GeO$_4$ compatible with both experimental magnetic susceptibility and ground state, thus resolving the discrepancies regarding the magnetic behavior of this compound. Several remarks are in order, though.

First, $J_1$ must be small, but no symmetry argument requires a complete cancellation of this coupling. The possible range of the $J_1$ values is determined by the energy difference between the UUDD and spiral states, as compared to the energy gain from the symmetric anisotropy. Quantum effects neglected within our LT analysis may also play a role here~\cite{zinke2009}. Detailed estimates go beyond the scope of our present manuscript but may be interesting if the symmetric anisotropy would be determined experimentally, e.g., by measuring magnon gap with electron spin resonance or THz spectroscopy.

Second, DFT proves incapable of estimating $J_1$ in a consistent manner (Table~\ref{tab:Cu2GeO4_ISO}). Similar problems were encountered for other short-range couplings in copper and vanadium compounds~\cite{tsirlin2010,tsirlin2011b}, although the Cu$_2$GeO$_4$ case appears to be most severe, because not only different flavors of DFT+$U$ but also different band-structure codes return largely different values of $J_1$. We attempted to vary the Coulomb repulsion $U_d$ and to change the double-counting correction, but were unable to reduce $|J_1|$ below 2\,meV using FPLO as the full-potential code. This indicates that a lot of caution should be taken in analyzing the short-range couplings obtained from DFT+$U$. On the more positive side, Cu$_2$GeO$_4$ may be an excellent test case for \textit{ab initio} methods, because the $J_1\simeq 0$ condition is very robust. Any significant deviation from it leads to the spiral order, which is not observed experimentally.

Third, spin canting caused by the DM interactions may give a clue to the formation of local electric polarization. The inverse DM mechanism triggers the polarization~\cite{katsura2005}
\begin{equation}
\mathbf{P}_{ij} \sim \boldsymbol{\epsilon}_{ij}  \times [\mathbf{S}_i \times \mathbf{S}_j],
\label{eq:Polarization}
\end{equation}
where $\boldsymbol{\epsilon}_{ij}=(0,1,0)$ is a vector connecting the magnetic sites $i$ and $j$ along the copper chains, and in a given layer the spins are presented by $\mathbf{S}_i = (\frac{1}{\sqrt{2}},\frac{1}{\sqrt{2}},\delta)$, $\mathbf{S}_j = (\frac{1}{\sqrt{2}},\frac{1}{\sqrt{2}},-\delta)$ for the up-up pair or $\mathbf{S}_i = (\frac{1}{\sqrt{2}},\frac{1}{\sqrt{2}},-\delta)$, $\mathbf{S}_j = (-\frac{1}{\sqrt{2}},-\frac{1}{\sqrt{2}},-\delta)$ for the up-down pair, where $\delta$ is small  canting.  Both pairs produce electric polarization of the same sign directed along $c$. This way, each copper layer generates a finite electric polarization that, however, cancels out between the neighboring layers following the symmetry of the $I_c\bar 42d$ magnetic space group~\cite{zou2016}. Nevertheless, it is conceivable that weak structural changes in the magnetically ordered state may reduce the symmetry, thus leading to a non-zero polarization. The step-like changes in the magnetic susceptibility and permittivity at $T_N$~\cite{zhao2018}, as well as the abrupt onset of the polarization in the magnetically ordered state~\cite{zhao2018,yanda2018}, may indicate a weak first-order nature of the magnetic transition, similar to $\alpha$-CaCr$_2$O$_4$, where the electric polarization has also been observed~\cite{singh2011} at odds with the symmetry of the magnetic structure~\cite{chapon2011}. On the experimental side, further thermodynamic measurements probing the nature of the magnetic transition in Cu$_2$GeO$_4$, as well as dielectric measurements probing the direction of the electric polarization, can be useful. 

In summary, we have shown that the collinear UUDD magnetic order in Cu$_2$GeO$_4$ is only possible in the $J_1\simeq 0$ limit and should be traced back to the nearly perfect compensation of the FM and AFM contributions to this exchange coupling. The UUDD order along the copper chains removes the frustration on half of the Cu$_4$ tetrahedra and, together with the weak symmetric anisotropy, leads to the peculiar magnetic structure with two different spin directions, as observed experimentally. 
%Weak spin canting allowed in this setting and triggered by the Dzyaloshinskii-Moriya interactions may be responsible for the ferroelectricity in the magnetically ordered state.

\acknowledgments
We would like to thank Yaroslav O. Kvashnin (Uppsala University)  for his help with the ELK code. The work of D.I.B. was funded by RFBR according to the research project No 18-32-00018. The work or V.V.M. was supported by the Russian Science Foundation, Grant No 18-12-00185.  A.A.T. acknowledges financial support by the Federal Ministry for Education and Research through the Sofja Kovalevskaya Award of Alexander von Humboldt Foundation.

\appendix
\section{ \label{app:Biquadratic}Biquadratic term in case of $S$ = 1/2}

Here we show that in the case of spin-$\frac12$ the biquadratic term $(\hat{\mathbf{S}}_{i} \hat{\mathbf{S}}_{j})^{2}$ can be re-written in the bilinear form $\hat{\mathbf{S}}_{i} \hat{\mathbf{S}}_{j}$. To this end, we use the property of the Pauli matrices,
\begin{eqnarray}
\hat{\sigma}^{a} \hat{\sigma}^{b} = \delta_{a b} \hat{I} + i \epsilon_{abc} \hat{\sigma}_{c},  
\end{eqnarray}
where $ \delta_{a b}$, $\hat{I}$, and $\epsilon_{abc}$ are the Kronecker delta, identity matrix, and Levi-Civita symbol, respectively.  Using the above relation and the commutation rule for two different sites, $[\hat{\sigma}_{i}^{a}, \hat{\sigma}_{j}^{b} ] = 0 $, it is straightforward to show that

\begin{eqnarray*}
(\hat{\mathbf{S}}_{i} \hat{\mathbf{S}}_{j})^{2} = \frac{1}{16} (\hat{\boldsymbol{\sigma}}_{i} \hat{\boldsymbol{\sigma}}_{j})^{2} =   \frac{1}{16} (\hat{\sigma}_{i}^{x} \hat{\sigma}_{j}^{x}  + \hat{\sigma}_{i}^{y} \hat{\sigma}_{j}^{y} + \hat{\sigma}_{i}^{z} \hat{\sigma}_{j}^{z} )^2 = \\ 
=  \frac{1}{16} ( \hat{\sigma}_{i}^{x} \hat{\sigma}_{i}^{x} \hat{\sigma}_{j}^{x} \hat{\sigma}_{j}^{x} +  \hat{\sigma}_{i}^{y} \hat{\sigma}_{i}^{y}  \hat{\sigma}_{j}^{y} \hat{\sigma}_{j}^{y} +   \hat{\sigma}_{i}^{z} \hat{\sigma}_{i}^{z}  \hat{\sigma}_{j}^{z} \hat{\sigma}_{j}^{z} +  \\  
+ 2 \hat{\sigma}_{i}^{x} \hat{\sigma}_{i}^{y} \hat{\sigma}_{j}^{x} \hat{\sigma}_{j}^{y} +  2\hat{\sigma}_{i}^{x} \hat{\sigma}_{i}^{z}  \hat{\sigma}_{j}^{x} \hat{\sigma}_{j}^{z} + 2  \hat{\sigma}_{i}^{y}  \hat{\sigma}_{i}^{z}  \hat{\sigma}_{j}^{y} \hat{\sigma}_{j}^{z} )  = \\ =  \frac{1}{16} ( 3 \hat{I} -  2 ( \hat{\sigma}_{i}^{z} \hat{\sigma}_{j}^{z} + \hat{\sigma}_{i}^{y}\hat{\sigma}_{j}^{y} + \hat{\sigma}_{i}^{x} \hat{\sigma}_{j}^{x})) =   \frac{3}{16} \hat{I} - \frac{1}{2} \hat{\mathbf{S}}_{i} \hat{\mathbf{S}}_{j}.          
\end{eqnarray*}

%\bibliography{main}

%

\end{document}